# Attenuation correction with Region Growing Method used in the Positron Emission Mammography Imaging System*

GU Xiao-Yue (顾笑悦)[1,2,1)] LI Lin (李琳)[1,2] YIN Peng-Fei (尹鹏飞)[1,2] Yun Ming-Kai (贠明凯)[1,2]
Chai Pei (柴培)[1,2] FAN Xin (樊馨)[1,2] HUANG Xian-Chao (黄先超)[1,2] SUN Xiao-Li (孙校丽)[1,2] WEI Long(魏龙)[1,2]

1 (Key Laboratory of Nuclear Radiation and Nuclear Energy Technology, Institute of High Energy Physics, Chinese Academy of Sciences, Beijing 100049, China)

2 (Beijing Engineering Research Center of Radiographic Techniques and Equipment, Beijing 100049, China)

**Abstract:** Positron Emission Mammography imaging system (PEMi) is a nuclear medicine diagnosis method dedicated for breast imaging. It provides a better resolution in detection of millimeter-sized breast tumors than whole-body PET. To address the requirement of semi-quantitative analysis with the radiotracer concentration map of the breast, a new attenuation correction method based on three-dimensional seeded region growing image segmentation (3DSRG-AC) solution was developed. The method gives a 3D connected region as the segmentation result instead of image slices. The continuously segmentation property makes this new method free of activity variation of breast tissues. Threshold value chosen is the key point for the segmentation process. The first valley of the grey level histogram of the reconstruction image is set as the lower threshold, which works fine in clinical application. Results show that attenuation correction for PEMi improves the image quality and the quantitative accuracy of radioactivity distribution determination. Attenuation correction also improves the probabilities to detect small and early breast tumors.

## 1 Introduction

Semi-quantification of vivo radiotracer is useful for nuclear medicine physicians to differentiate between benign and malignant tissues. Positron emission tomography (PET) has a reputation for semi-quantitative analysis with the accurate and precise radiotracer concentration map of the whole body. However, the exact activity map is not so easy to achieve and there are many factors to effect, such as photon attenuation, count-rate loss due to dead time, variations in detector efficiency et al. [1]. Among them, most effects have been well corrected during the development of PET technology. For example, a generic protocol for attenuation corrections (AC) is combined PET with a transmission scan, such as a rod source or computed tomography (CT) [2-3]. Nevertheless, attenuation correction is still a challenge in a transmission-less whole-body PET system. Since the whole-body PET detects a complicated and comprehensive radiotracer distribution image consisting of bone, air, soft tissue and other components, it makes the precise image segmentation from activity map difficult and challenging.

In recent years there is a boom in developing organ-dedicated nuclear medicine techniques. Among the systems, positron emission mammography has achieved the fastest development. With a small and compact detector, positron emission mammography has a better resolution and image quality [4-6]. A new polygon Positron Emission Mammography imaging system called PEMi with the ability to detect millimeter-sized lesions had been developed in 2009 by Institute of High Energy Physics, Chinese Academy of Sciences. It is capable of producing better quality breast-PET images compared to standard methods [7]. PEMi is without transmission scanning after an overall consideration. To achieve semi-quantitative analysis in PEMi system, a generic protocol for AC is using image segmentation method. What's more, the radiotracer distribution of the breast is clear and simple, which makes AC based on image segmentation easy and practice in PEMi.

AC methods in transmission-less PET systems have been studied for many years. Early approaches about AC methods starting with the pioneering work by Censor et al. used iterative algorithm to extract activity map and attenuation map out of emission data simultaneously [8]. But the major limitation of the inherent cross-talk between the two maps made it unpractical for clinical applications.

Received date 15 December 2014
*Supported by Knowledge Innovation Project of Chinese Academy of Sciences (KJCX2-EW-N06)
1) E-mail: guxiaoyue@ihep.ac.cn

Natterer et al. proposed a more popular approach to estimate the attenuation map based on the data consistency conditions (DDCs) [9]. DDCs method had been tested in the PEMi system already. It was powerful in most situations, but if the activity distribution of the cancer lesion was too concentrated than the breast tissues, it would only give the breast tumors as the segmentation result and was impossible to segment between the breast tissue and the background regions [10]. Manual boundary modification is easy to realize but the major disadvantage is time consuming for application.

The well-known seeded region growing (SRG) algorithm is a powerful and flexible approach to image segmentation [11]. It is well used in search of homogeneous regions inside the image based on connectivity and similarity properties among the voxels. The basic process is starting from an initial set of voxels, known as "seeds". Then the nearby voxels are selected according to some predefined criteria. The advantages of the method are robust, rapid, and free of tuning parameters while the disadvantage is sensitive to noise. The development of SRG method in three-dimensional (3D) had already been studied and applied on other clinical imaging methods, such as CT and MRI [12-14].

In this study, a method was proposed to perform the attenuation correction with 3D seeded region growing (3DSRG-AC) image segmentation algorithm. 3DSRG method gives a 3D connected region as the segmentation result instead of two-dimensional (2D) image slices. The 3D connective property makes the method free of concentration variations of breast tumors, and all the breast tissues consisting of both normal tissues and lesions are segmented as a whole. The 3DSRG-AC method based on a proper threshold between breast tissues and the air region works fine in the differentiate process. The key point of this method is a prior gaussian filter application and set the first valley of the grey level histogram of the image as the lower threshold of the region growing method. We had tested the method with the experimental data and the clinical data on PEMi which have achieved stable and robust results. The dedicated breast PET was made of Institute of High Energy Physics, Chinese Academy of Sciences [7].

## 2 Methods
### 2.1 System Design

The detector of PEMi is designed in a polygon structure, which is constructed with 16 modules. These modules link in the shape of a detector ring, as shown in Fig. 1(a). Each module consists of four blocks, and each block is arranged in 16 × 16 crystal arrays with a pixel size of 1.9 mm × 1.9 mm × 15 mm. The crystals are made of cerium-doped LYSO. To improve light transmission, the top surface of each crystal is roughened, and the other sides (the four sides around and the side coupled to the PSPMT) are optically polished. An enhanced specular reflector is used between the crystal elements to reduce optical crosstalk.

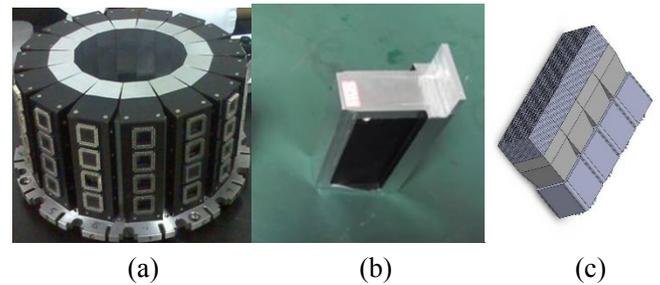

(a)　　　　　　(b)　　　　　　(c)

Fig. 1. Detector design. (a) Photograph of the PEMi detector ring. (b) Photograph of a PEMi module. (c) Schematic of the PEMi block.

A position-sensitive photomultiplier tube (PS-PMT, Hamamatsu R8900-C12) is coupled to the end face of the scintillator arrays with a tapered light guide, as shown in Fig. 1(c). The specially designed light guide is made of optical fiber to guarantee high light transmission, thus linking the scintillator array (32 mm × 32 mm) and the PMT (23.5 mm × 23.5 mm). The module design is shown in Fig. 1(b). The pixel map of one detector has been obtained (see Fig. 2a). All the 16 × 16 crystals are clearly identifiable in the flood histogram. And the associated profiles also shown in the figure (see Fig. 2b)

Analog signals from the PSPMT are first processed in the front-end electronics. The resistor chains reduce the 6 (X) + 6 (Y) anode signals from each PSPMT into four channel signals. The timing signals are acquired from the PMT's last dynode, processed by the constant fraction discriminator, and then transmitted to the data acquisition electronics system together with the processed analog signals.

The axial FOV spans 128 mm, whereas the transaxial FOV is restricted to a diameter of 110 mm by the software after list-mode acquisition. The system operates in full 3D acquisition mode, permitting coincidences between any two detector rings. Both the list mode and histogram mode are provided in the acquisition process. In the histogram mode, coincidence event data are sorted into sinograms and converted into 2D slice data by single slice rebinning (SSRB) and Fourier rebinning (FORE) algorithms using different span and ring differences. Images are reconstructed with a pixel size of 0.5 mm × 0.5 mm in a matrix of 256 × 256. The 2D filtered back projection (FBP) algorithm or the 2D ordered subsets expectation maximization (OSEM) algorithm can be chosen through user interface. The 3D Gaussian post-filter is optional in the reconstruction process to reduce image noise. The reconstructed images are then displayed in 3D mode and the sagittal, coronal, and transverse views are provided. [7]

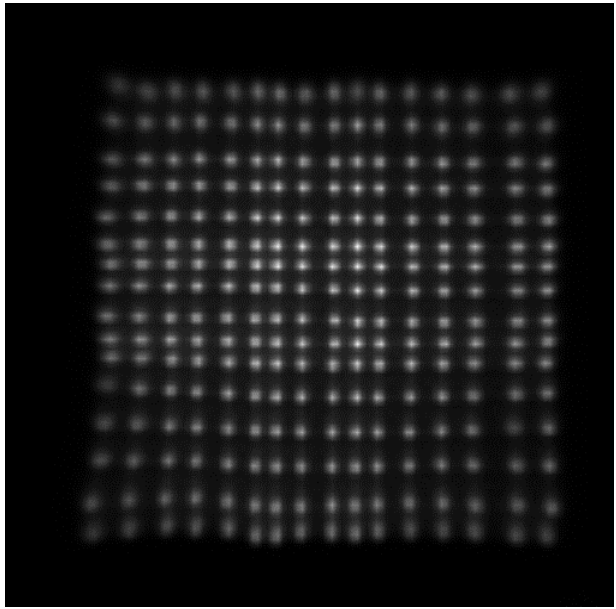

(a)

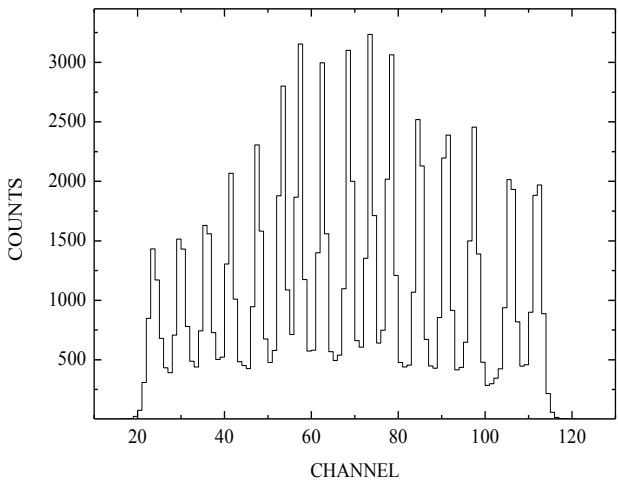

(b)

Fig. 2. Detector performance. (a)A pixel map from one detector's arrays shows good detector element separation. (b)Profiles of center channel of the flood histogram.

## 2.2 Attenuation correction with Region Growing Method

Flow char of 3DSRG-AC algorithm is presented in Fig. 3. First time reconstruction of sinogram is performed before image segmentation. And then the reconstructed PEMi image is segmented by 3DSRG method. Siddon ray tracer projection is processed on the segmentation result as follow to obtain AC coefficients for each line of coincidences (LOR) in sinogram. In addition, the LORs in sinogram are multiplied with their AC coefficients. After procedures mentioned above, AC has already been performed in PEMi. But to obtain the attenuation corrected PEMi image, the image reconstruction process has to been performed again.

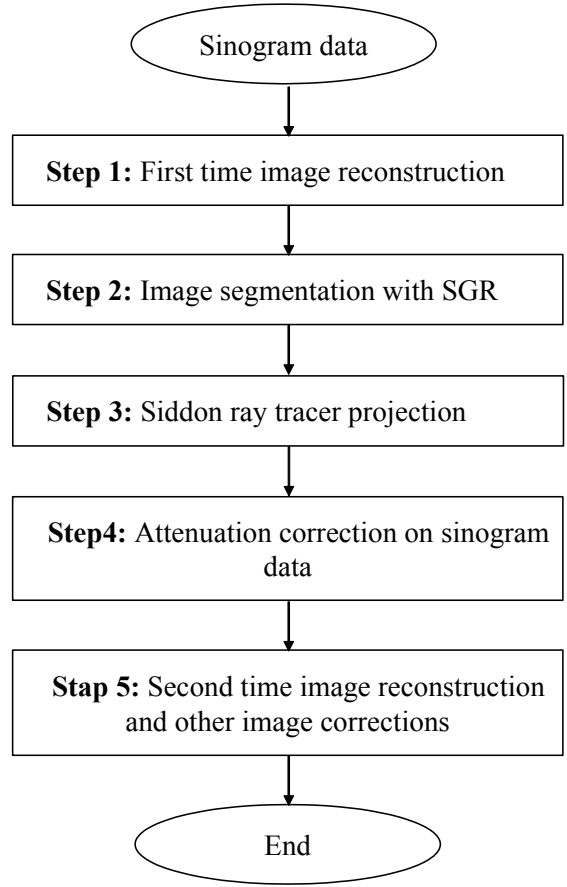

Fig. 3. Flow chat of the 3DSRG-AC algorithm.

The key procedure of 3DSRG-AC algorithm is segmentation step. It is starting from an initial set of voxels, known as "seeds". Then the nearby voxels are grouped into the growing region when they are satisfied the predefined criteria. Whenever a new voxel involved in the growing region, a new round traverse over its neighborhoods started. The iterative process stops until all voxels have been traversed or no voxel meets the predefined criteria. The predefined parameters are set as follows in 3DSRG-AC:

**Seeds:** Max value of the PEMi image is chosen as the growing seed for the algorithm;

**Growing criteria:** The lower threshold of the interval is chosen as the first valley of the grey level histogram after a Gaussian filter application for the PEMi image, see the red triangle mark in Fig.4; the upper threshold is chosen as the max value of the image;

**Stop criteria:** no voxel meets the growing criteria;

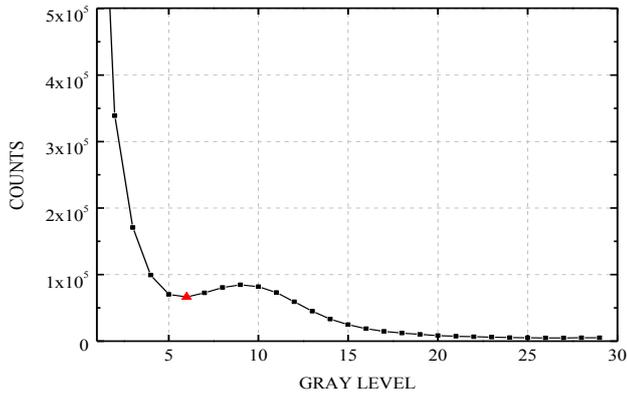

Fig. 4. Grey level histogram of the PEMi image. The red triangle mark show the first valley of the grey level histogram of the PEMi image, which is set as the lower threshold.

3DSRG-AC method choses 16 neighborhood for the 3D growing processes, which is effective and fast, as shown in Fig. 5. The breast tissues are separated from the air based on the specified threshold in step two. Then different attenuation coefficients are assigned to breast and air regions respectively. For the breast and air, the coefficients are as follows [15]:

$$\mu_{breast} = 0.0096 \, mm^{-1}, \qquad (1)$$
$$\mu_{air} = 1.04 \times 10^{-5} mm^{-1}. \qquad (2)$$

To get the attenuation matrix, the PEMi image values are reset accord with the segment result. Each LOR get the attenuation coefficient from the attenuation matrix projection. In step three Siddon ray tracer is applied for the projection process [16]. The reconstruction process is performed again with the attenuation corrected LORs. The LOR coefficients are performed as follows:

$$AC_{LOR} = exp^{(\mu_{breast} \cdot L_{breast})} \cdot exp^{(\mu_{air} \cdot L_{air})}. \qquad (3)$$

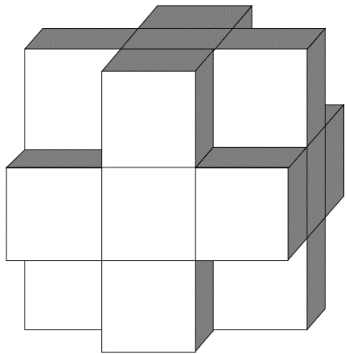

Fig. 5. Neighborhoods for 3D growing process.

## 3 Results:
### 3.1 Simulation Experiments

We had tested 3DSRG-AC method by monte carlo simulations. Two cylindrical phantoms both with external diameter of 100 mm and axial height of 157 mm were employed. One of the phantoms was filled with uniform 18F-FDG solution having activity concentration of 200nCi/cm3. The other phantom is similar with the one above but add extra four small sphere sources with different diameters, respectively 4mm, 3mm, 2mm and 1.5mm. All the PEMi data were reconstructed with OSEM algorithm with 2 iterations and 8 subsets. The attenuation corrected and the uncorrected pictures were both performed with a Gaussian filter.

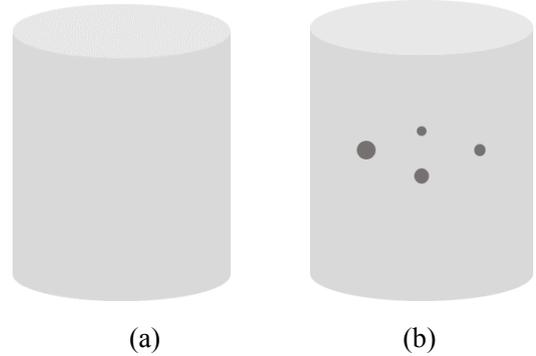

Fig. 6. Pictures of phantoms. (a) Uniform phantom. (b) Point Source phantom.

### 3.1.1 Uniform phantom

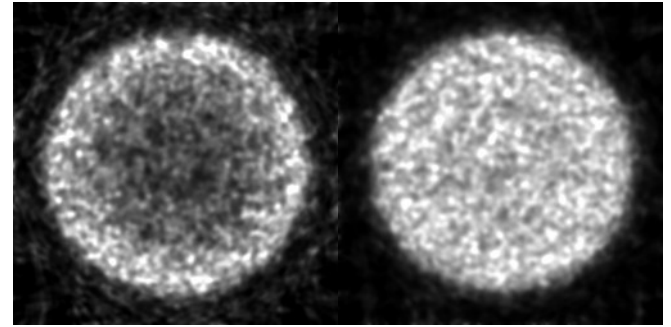

Fig. 7. The reconstructed result of uncorrected view and corrected view. (a) Corrected result. (b) Uncorrected result.

Fig. 7 shows the reconstruction results of the uniform phantom. The attenuation corrected image is more uniform than the uncorrected one. Activity of the uncorrected image declines from the phantom margin to the center. For a detail quantitative analysis of the results, annular regions of interest (ROI) with 10 mm width were analyzed, as shown in Fig. 8(a). Quantitative analysis of the results are shown in Fig. 8(b).

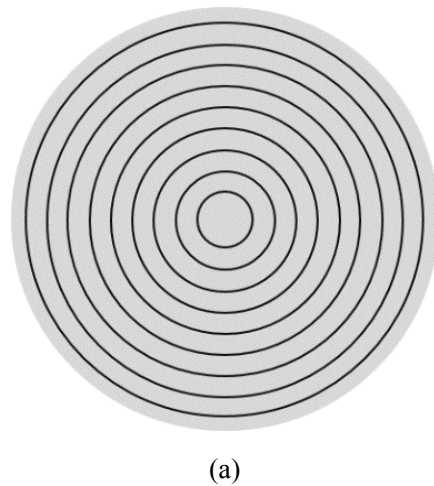

(a)

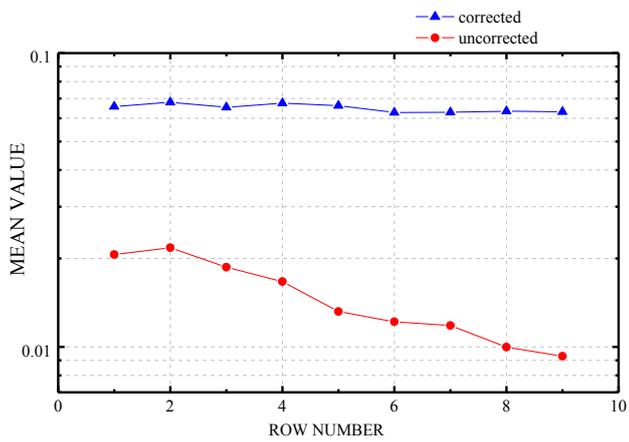

(b)

Fig. 8. Quantitative analysis of uncorrected view and corrected view. (a) Schematic diagram of the ROIs. (b)Quantitative analysis result of the uncorrected and corrected views.

### 3.1.2 Point source Phantom

With the same geometry, the uniform phantom added with 4 point sources also had been tested. The reconstructed and attenuation correction methods were as same as the uniform phantom. The maximum intensity projection (MIP) view and transviews of the results are showed in Fig. 9. Uncorrected image may render a wrong result for the un-uniform activity distribution. But the corrected image shows a great contrast for the hot sources site.

### 3.2 Clinical Result

Clinical data were tested as followed. The three views of segmentation result are shown in Fig.10. Attenuation results are given in a MIP view in Fig.11. Decay, dead time, random, normalization and scatter corrections were all performed for the clinical image. The corrected PEMi data show two lesions of the breast, which were confirmed as an intra-ductal carcinoma by biopsy. There were several suspicious abnormality lesions exited in the case, as the red arrows in Fig. 11 clearly shows. The corrected picture show more information than the uncorrected one, and has a better contrast.

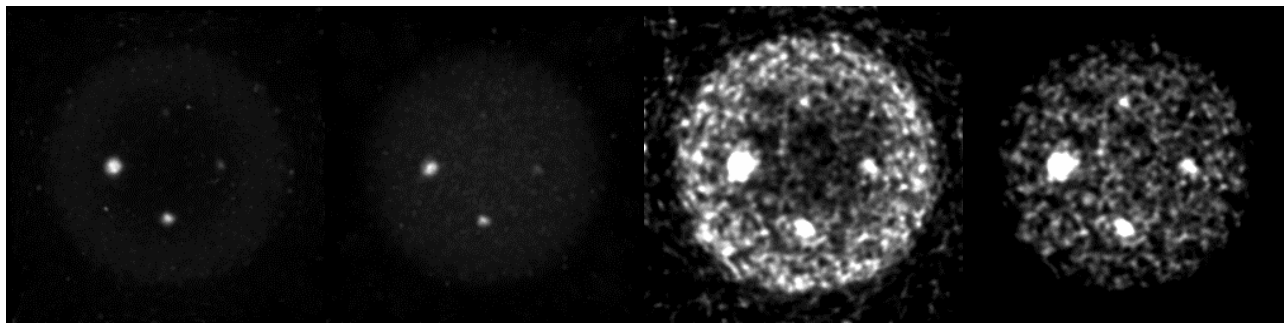

(a)      (b)      (c)      (d)

Fig. 9. MIP views and transviews of the reconstructed uncorrected and corrected pictures of point source phantom. (a) MIP view of the uncorrected result. (b) MIP view of the corrected result. (c) Transaxial view of the uncorrected result. (d) Transaxial view of the corrected result.

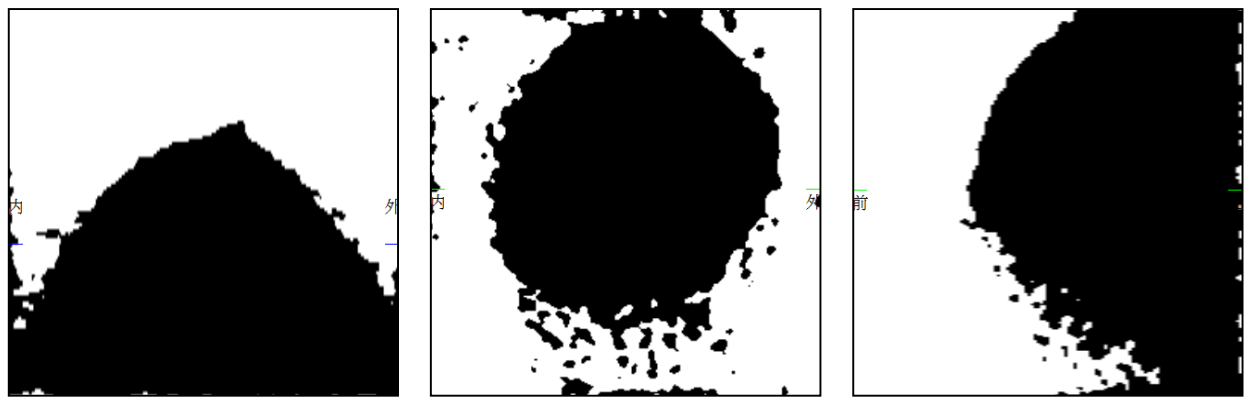

(a)      (b)      (c)

Fig. 10. Segment results of clinical data. (a) Transverse view of segment result of clinical data. (b)Coronal view of segment result of clinical data. (c) Sagittal view of segment result of clinical data.

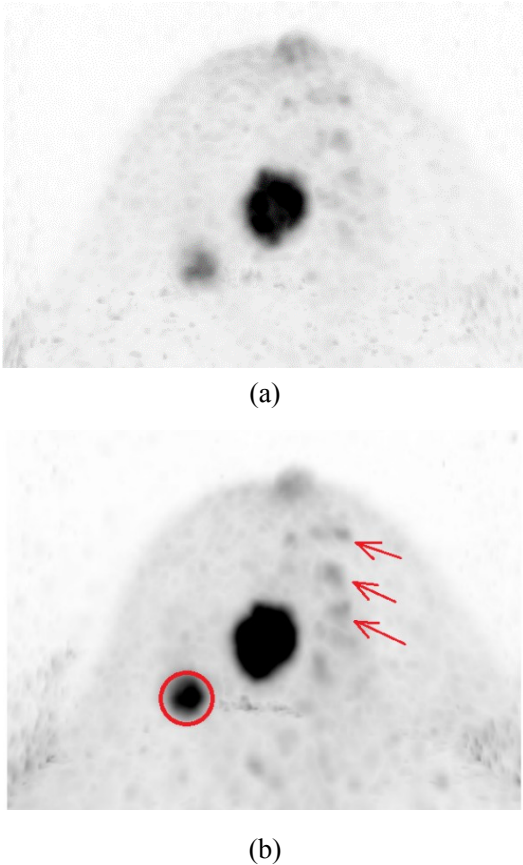

(a)

(b)

Fig. 11. Reconstruction result of clinical data. (a) Uncorrected MIP view shows two main lesions. (b) The small lesion in the corrected MIP view is more clearly than the uncorrected view, as same as the several suspicious abnormality lesions in the case.

## 4 Discussion

We developed an attenuation correction method based on 3D region growing algorithm, namely 3DSRG-AC. The most advantage of 3DSRG-AC method is that it gives a connected region as a segmentation result instead of 2D image slices. This 3D segmentation property takes the advantage of consecution of the activity distribution in breast tissues, which makes the algorithm free of concentration variations of breast tumors. As a result, the breast tissues consisting both normal tissues and lesions are segmented as a whole based on their gray level distribution. The slice by slice 2D segmentation method easily leads to inconsistent results between slices, because the different activity distribution between normal tissues and lesions. That's say, when the lesions is too concentrated, 2D segmentation method easily take the lesion boundary as the contour and may render a mistake.

3DSRG-AC method ensures a stable and accurate segmentation result. Threshold value chosen is the key point for the segmentation process. Inappropriate threshold value chosen will influence attenuation correction effect. For example, the first valley of the grey level histogram of the point source phantom was tested as 10, which was the optimal estimated value of the threshold between breast tissues and air, see the red triangle mark in Fig. 12. The proper threshold value was set as the lower threshold to perform AC which got a proper result, see Fig. 13(a~c). When the half value of the proper threshold was tested as the new segment parameter, which had been located in the air domain (see the blue diamond mark in Fig. 12), the tested results showed burrs on contour edges due to the noise interference as well as the smallest hot point source also couldn't been distinguished for the excessive attenuation correction, see Fig.13.

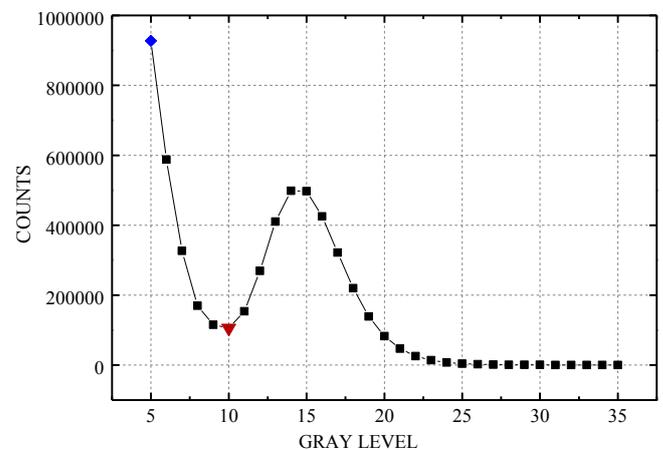

Fig. 12. Grey level histogram of the PEMi image. The red triangle mark show the optimal threshold. The blue diamond mark show the half value of the proper threshold

A lower threshold adds noise information in the segmentation image. A more aggressive attenuation coefficient contributed from extra noise information in segmentation result will further reduce noise interference as well as other activity information. Therefore, we confirm that an appropriate segmentation shape adapt for real phantom or breast tissues is accurate, while excessive attenuation correction will result in the loss of useful information. In the practice, we use the first valley of the grey level histogram as the lower threshold, which works fine in the clinical application.

As a dedicated-organ imaging system, PEMi is more compact than whole body PET. And the gamma rays decay is less than the whole body PET. But the phantom data and clinical data show that attenuation correction for PEMi is still important. Fig. 8(b) shows the activity profiles drop as depth in the uniform phantom increases in uncorrected image. The uniform phantom achieved a more uniform result after attenuation correction. This effect will affect small lesion detection in the marginal area. As shown in

Fig. 9, it's a little difficult to distinguish the smallest source from the noise, while the small site can be distinguished in the AC performed image.

Attenuation correction effect in clinical data is shown in Fig .11. The two views corrected and uncorrected both indicated there were two lesions in the breast, which had been confirmed as an intra-ductal carcinoma by biopsy. The small lesion was more obvious in the corrected view than the uncorrected view, as the red circle shows in Fig.11(b). Further, there were several suspicious abnormality lesions exited in the case, as the corrected view clearly shows while the uncorrected view is unclear, see the red arraw marks in Fig.11(b). The corrected picture shows more information than the uncorrected one, and has a better contrast. We confirm that attenuation correction is important in the clinical application for PEMi.

## 5 Conclusion

PEMi system is a nuclear medicine diagnosis method dedicated for breast imaging. It provides a better resolution in detection of millimeter-sized breast tumors. With a spatial resolution of approximately 2 mm, the PEMi system is capable of producing better quality breast-PET images compared to other nuclear imaging methods. The AC method is based on 3D region growing algorithm, namely 3DSRG-AC. The most advantage of 3DSRG-AC method is that it gives a connected region as a segmentation result instead of 2D image slices. This 3D segmentation property takes the advantage of consecution of the breast tissues, which makes the algorithm free of concentration variations of breast tumors. The method ensures a stable and accurate segmentation result. 3DSRG-AC also improves the probabilities to detect small and early breast tumors. Results show that attenuation correction for PEMi improves the image quality and the quantitative accuracy of radioactivity distribution determination.

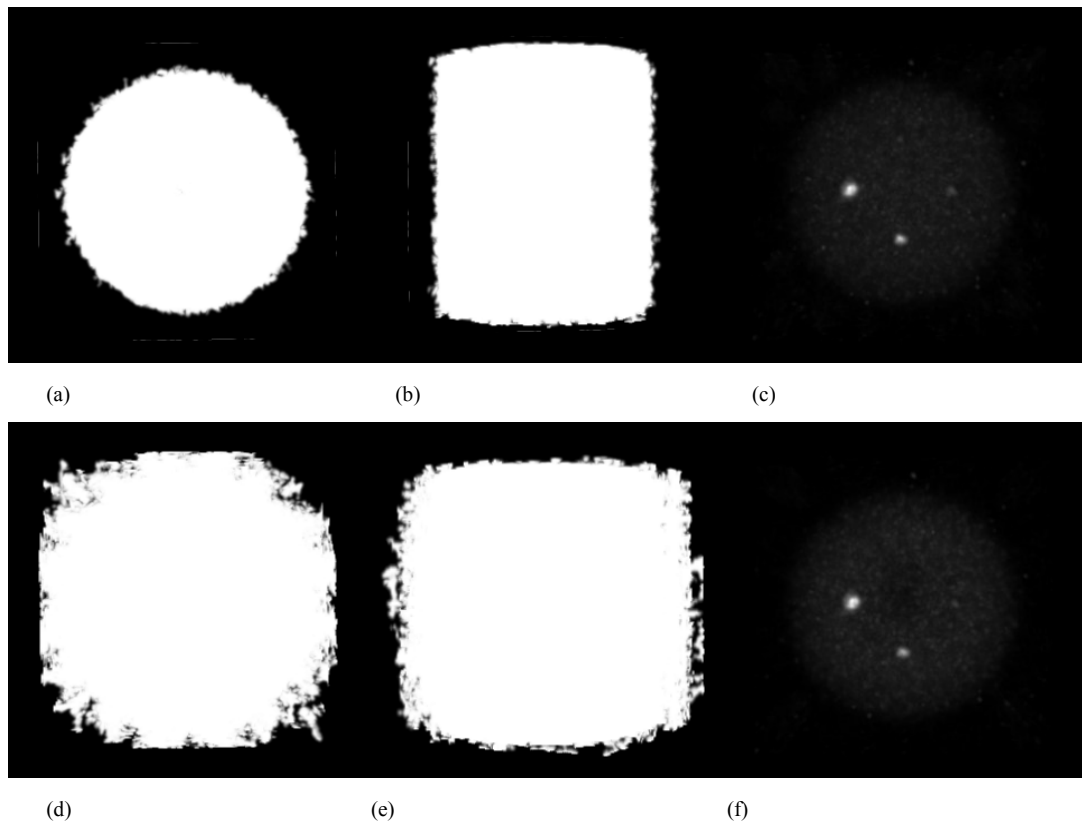

Fig. 13. Different segmentation thresholds comparison. (a) Transview of the proper threshold segmentation result. (b)Sagittal view of the proper threshold segmentation result. (c) Reconstructed result after AC of the proper threshold segmentation result. (d) Transview of the small threshold segmentation result. (e)Sagittal view of the small threshold segmentation result. (e) Reconstructed result after AC of the small threshold segmentation result.